\documentclass{amsart}
\usepackage[utf8]{inputenc}
\usepackage{mathtools}
\usepackage{physics}
\usepackage{makecell}
\usepackage[textwidth=14cm]{geometry}
\usepackage[backend=biber, sorting=none, style=nature, citestyle=ieee, url=false, isbn=false]{biblatex}
\usepackage{graphicx}

\usepackage{lineno}

\AtEveryBibitem{
  \clearfield{day}
  \clearfield{month}
  \clearfield{issue}
}

\begin{document}

\title{3D Zeros in Electromagnetic Fields}
\maketitle
\author{Alex J. Vernon$^{1\dagger}$, Mark R. Dennis$^{2\ddagger}$, and Francisco J. Rodr\'iguez-Fortu\~no$^{1*}$\\}

\address{$^1$Department of Physics and London Centre for Nanotechnology, King's College London, Strand, London WC2R 2LS, UK \\\indent$^2$School of Physics and Astronomy, University of Birmingham, Birmingham B15 2TT, UK\\}

\email{$^{\dagger}$alexander.vernon@kcl.ac.uk\\ \indent$^{\ddagger}$m.r.dennis@bham.ac.uk\\ \indent$^{*}$francisco.rodriguez\_fortuno@kcl.ac.uk}

\begin{abstract}
We present a study of 3D electromagnetic field zeros, uncovering their remarkable characteristic features and propose a classifying framework. 
These are a special case of general dark spots in optical fields, which sculpt light's spatial structure into matter-moving, information-rich vortices, escape the diffraction limit for single-molecule imaging, and can trap particles for nanoscale manipulation. 
Conventional dark spots are two-dimensional in two aspects: localised in a plane and having a non-zero out-of-plane field component. 
We focus on non-paraxial fields, where three-dimensional dark spots can exist non-stably at fully localised points, making distinct imprints in the flux of energy and momentum, and in the light’s polarisation texture. 
With this work, we hope to enhance current dark spot applications, or inspire new ones impossible with lower-dimensional zeros.
\end{abstract}

\section{Introduction}

\noindent An optical vortex is the name commonly given to a zero in a complex scalar field, such as a component of the electric $\mathbf{E}$ or magnetic $\mathbf{H}$ field. 
Vortices in these components occur naturally in general 3D monochromatic interference \cite{OHolleran2009}, where they are infinitely thin continuous strands either extending infinitely through space, or coiled into knotted, un-knotted or linked closed loops \cite{Leach2004,OHolleran2006,Dennis2010,Tempone2016}. 
On a vortex strand, the phase of the complex scalar field (with zero real and imaginary parts) is undefined creating circulation in the phase of the rest of the field. 
This phase increases in a clockwise or anti-clockwise sense by an integer multiple of $2\pi$ along any closed loop containing one vortex line. 
Vortex lines in optics have direct analogues in acoustics and water waves, and as a type of topological defect, are related to vortices in (super)fluids \cite{Kleckner2013} and in Bose-Einstein condensates \cite{Weiler2008}, and even cosmic strings \cite{Hindmarsh1995}. 
Strong research interest in optical vortices over the past 30 years, combined with the availability of instruments and the flexibility in generating \cite{Guo2020,Wang2019,Li2021,Zhang2016} and structuring \cite{Lim2021} vortex-carrying beams, has positioned optics to act as a sandbox for exploring topological phenomena that appear more broadly across physics.\\
\indent When considering the full 3D vector characteristics of an optical field, vortex lines in individual field components like $E_x$, $E_y$, and $E_z$ are basis-dependent and not so physically meaningful. 
By picturing these different scalar vortex threads permeating the vector field, we can appreciate how unlikely it is that the optical field is zero at a point (i.e.~$\mathbf{E=0}$, all three components simultaneously zero) in typical 3D interference (the vortex line in each of the three field components would meet at such a zero point, requiring the manipulation of three extra parameters beyond the spatial $x,y,z$). 
Despite the rarity of zeros in the wild, a lower-dimensional version can be readily manufactured in optical beams, and is remarkably well-studied. 
Paraxial doughnut beams have an axial zero in the transverse field surrounded by a bright ring, and are used in modern spectroscopy techniques \cite{Balzarotti2017,Hell1994} because of the zero's immunity to the diffraction limit. 
The transverse field effectively consists of one or two scalar components with the vortex line along the beam axis, causing the real part of the local wavevector to curl around the axis and imbue the beam with intrinsic orbital angular momentum. 
The longitudinal field, meanwhile, is non-zero (albeit very small due to paraxiality) in the centre of the beam which, therefore, is better imagined not as an exact axial zero, but as a dim line of linear polarisation (an L line) polarised parallel to the beam direction. 
This, and its confinement in only two dimensions, stretching along the third, is why we refer to the almost-dark centre of the doughnut beam as a two-dimensional zero. 
Its topological index is straightforward to define by counting how many multiples of $2\pi$ the phases of the transverse components climb through over an enclosing circuit. 
The intrinsic orbital angular momentum carried by doughnut beams is the key property of the spatial structure of light that can rotate matter \cite{He19951,He19952} and store information \cite{Wang2012,Huang2014,Willner2021}.\\
\indent Surprisingly, the fully localised, three-dimensional optical field zero, $\mathbf{E=0}$, has been left largely unexplored. 
This is probably due to its unstable nature---a perturbation will destroy the zero point (i.e.~cause the vortices in the three components no longer to coincide).
Nevertheless, such a point is theoretically possible and can be artificially synthesised \cite{Vernon2022}, but very little is understood about how it is imprinted into the surrounding field, and there is no classifying topological index like the topological charge of a 2D vortex. 
The 3D electromagnetic field zero is the focus of this work. 
A zero in the $\mathbf{E}$-field alone has codimension 6, requiring that the six total degrees of freedom of two real, three-dimensional vectors (the real and imaginary parts of the three components $\mathbf{E}$) are suppressed simultaneously.
This means 3D zeros exist stably in a six-dimensional parameter space, and is why optical field zeros are not natural in random interference patterns spanning only three spatial dimensions, being hidden by instability.
Instead, 3D zeros must be revealed by tuning an additional three parameters (this is discussed in \cite{Spaegele2022} for a zero in two electric field components).
Some of these parameters could be the polarisation components of a plane wave, for example, and in fact, 3D zeros can be very easily manufactured and controlled in pure plane wave interference or near fields with a simple technique \cite{Vernon2022}, and their higher dimensional confinement could provide a greater degree of precision in dark spot spectroscopy.
Due to their electric field dependence, the zero in $\mathbf{E}$ is coupled to a collection of singularities, each with its own topological signature, in various physical quantities associated with the light field including the complex Poynting vector, canonical momentum, spin momentum and spin angular momentum. 
Learning how energy flow and momentum circulate around a 3D vortex could inspire applications which would be otherwise unfeasible using typical lower dimensional zeros. 
Alternatively, the magnetic field $\mathbf{H}$ may vanish at a point, or more extremely, both $\mathbf{E}$ and $\mathbf{H}$ might simultaneously vanish, giving a true electromagnetic null with codimension 12.
Here, we report the key features of a 3D field electric or magnetic field zero, including the way that polarisation singularities are forced to intersect and the flux of the complex Poynting vector and canonical and spin momentum. 
With these findings, for the first time, we propose a framework to classify the physically realisable varieties of 3D field zero.

\section{Results}

To contextualise our study, we begin with some brief intuition on the special features which we might expect to find near to a 3D zero.\\
\indent If either $\mathbf{E}$ or $\mathbf{H}$ is zero at a point $\mathbf{r}_0$, then of that field, say $\mathbf{E}$, the flux of energy, canonical momentum, spin angular momentum (and other quantities) are zero too. 
Since these fluxes are vector quantities, their direction is singular at $\mathbf{r}_0$ and an imprint is made in the surrounding space where they are well-defined. 
In three spatial dimensions, even if these fluxes are divergence-less, there is more than one possible (topologically unique) imprint which can be left by and characterise the zero in $\mathbf{E}$. 
The electric field spin is particularly interesting, because its zeros (in non-paraxial fields) are co-dimension 2 objects---meaning they are one-dimensional continuous lines, defining the threads of pure linear electric polarisation. 
This continuity should require at least one zero-spin line, an L line, to pass through $\mathbf{r}_0$. 
A similar argument can be made for lines of pure circular electric polarisation, except that C lines are defined by a complex quadratic equation, $\mathbf{E\cdot E}=0$, equivalent to a real quartic equation, $|\mathbf{E\cdot E}|^2=0$, which has either zero, two or four real roots. 
It turns out, as we will show, that a given number of C lines and L lines must always intersect in a 3D electric field zero. 
Before reporting these and other findings in detail from mathematical argument and analytical simulations in section 2.3 and beyond, the next two subsections 2.1 and 2.2 provide an overview of polarisation singularities and set out our way of classifying 3D field zeros using dyadics associated with the field.

\subsection{Overview of Polarisation Singularities in Paraxial and Non-Paraxial Fields}

\begin{figure}
    \centering
    \includegraphics[width=\textwidth]{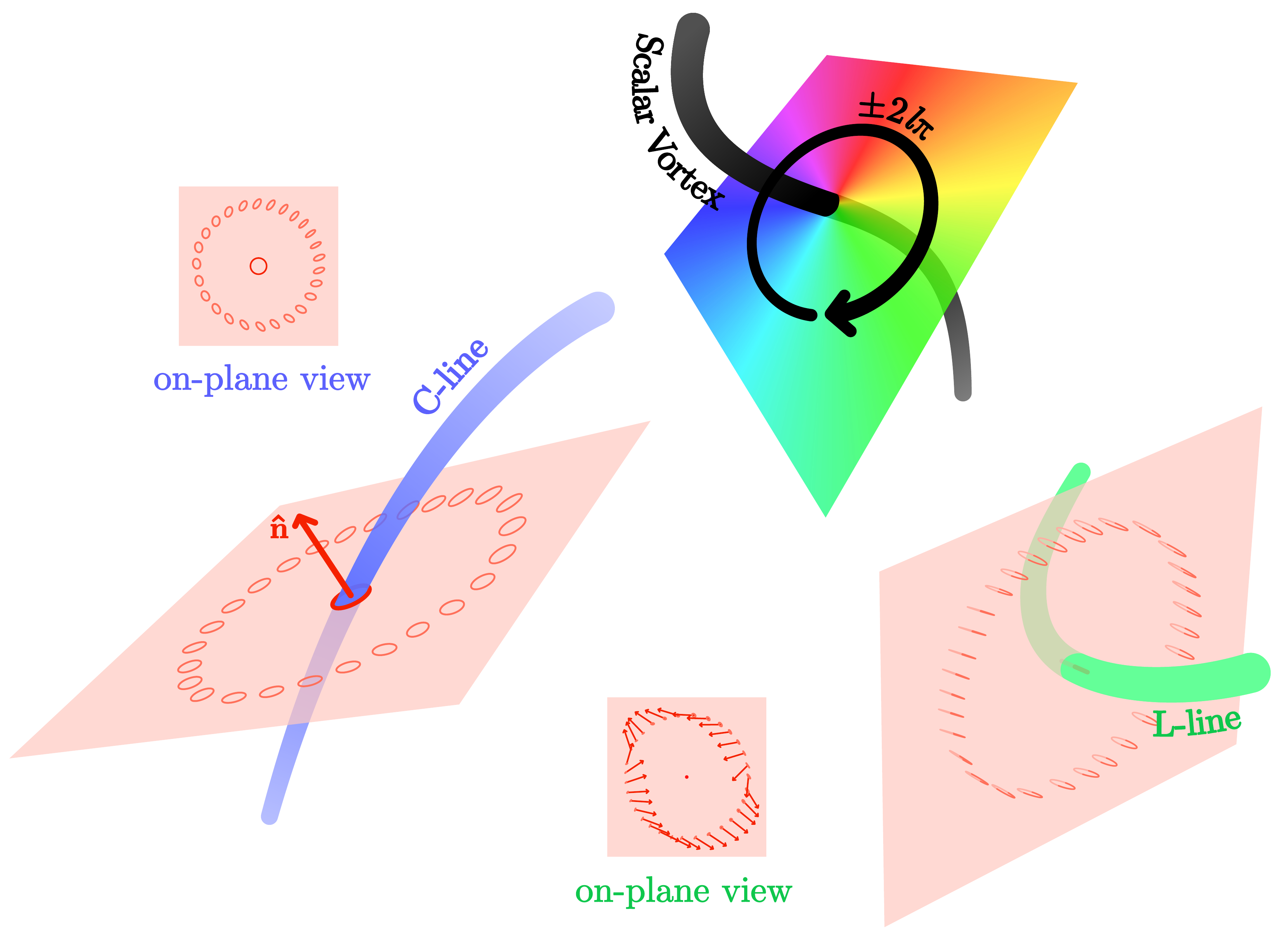}
    \caption{Visualisation of scalar and polarisation singularities in a non-paraxial electromagnetic field. 
    Scalar vortices (black line) exist in complex scalar fields, such as the components of $\mathbf{E}$, where the scalar field is zero and its phase is undefined, forming 1D threads in the interference of three or more plane waves. 
    Around a scalar vortex line, the phase of the field increases by an integer $l$ multiple of $2\pi$ in a clockwise or anticlockwise sense. 
    Singular lines exist in the complex vector characteristic of $\mathbf{E}$ and $\mathbf{H}$ fields, called polarisation singularities, which include C lines (lines of circular polarisation) and L lines (lines of linear polarisation). 
    In a circuit around a point on a C line (blue line), in the plane of the polarisation circle at that point, nearby polarisation ellipses rotate through an integer multiple of $\pi$ radians. 
    Around an L line (green line), the normal to nearby polarisation ellipses rotates by an integer multiple of $2\pi$ radians.}
    \label{fig:explain}
\end{figure}

\indent L lines and C lines are called polarisation singularities and are the vector version of scalar vortex lines in wave fields, existing in light \cite{Nye1987,Nye1983,Berry2001}, acoustic and water waves \cite{Bliokh2021} (both acoustic and water waves have a vector nature \cite{Bliokh2019,Bliokh2022}) where some property of the general polarisation ellipse is not defined. 
In 3D fields, polarisation singularities are often described as the underlying skeleton which embeds highly complex topologies into the field's polarisation texture \cite{Sugic2021,Larocque2018}. 
Polarisation singularities have been studied in full 3D and in paraxial fields \cite{Dennis2002}, where in paraxial fields (considering only the two transverse field components), polarisation is circular at points and linear along lines. 
Propagating the paraxial field (maintaining the transverse polarisation) draws out the C points and L lines in the transverse plane into C lines and L surfaces in three dimensions.\\
\indent A polarisation ellipse has orthogonal semi-major and semi-minor axes, telling us which way the ellipse is oriented. 
But because a polarisation circle has no semi-major or semi-minor axes, at a C point, the orientation of the circle is undefined causing neighbouring polarisation ellipses (almost circular) to rotate when tracked along a C point-enclosing loop. 
The ellipse major axis is described throughout space with a line field, in that the axis is oriented some way in space but does not point one way or another---an ellipse looks identical to its 180 degree rotated self. 
This means that along the enclosing circuit, the rotating ellipses turn continuously through an integer multiple of $\pi$ radians, rather than $2\pi$, which is why C points are assigned a half-integer index. 
When the field is fully three dimensional and the polarisation ellipse is free to tilt in any Cartesian direction, circular polarisation still occurs along one-dimensional threads (C lines which are no longer straight as in the paraxial case) but the surrounding polarisation ellipses also twist, so that their major axes sweep out M\"obius strips \cite{Freund2010,Dennis2011,Bauer2015}. 
Analogues of C lines exist in polychromatic fields, shaping the rest of the field into other remarkable topological structures \cite{Pisanty2019}.\\
\indent L lines/L surfaces in paraxial fields (ignoring longitudinal fields) separate regions of left and right handed polarisation ellipses. 
In non-paraxial fields, L lines are strictly one-dimensional lines (not surfaces) and complement C lines in shaping the surrounding polarisation structure. 
This reduction of dimension to the L entity occurs because, to be linearly polarised, the real and imaginary parts of the field (say $\mathbf{E}=\mathbf{p}+i\mathbf{q}$) need to be (anti)parallel (not necessarily equal). 
If $\mathbf{E}$ is paraxial and linearly polarised, then in the transverse plane, the ratio of the $x$ components of $\mathbf{p}$ and $\mathbf{q}$ must equal the ratio of their $y$ components---a single condition, dissolving only one degree of freedom of one vector relative to the other. 
If $\mathbf{E}$ is non-paraxial, then an extra condition accounting for the ratio of the $z$ components of $\mathbf{p}$ and $\mathbf{q}$ must be satisfied for linear polarisation \cite{Nye1987}. 
Between paraxial and full 3D fields, the linear polarisation object's codimension, which is the dimension of the electric spin angular momentum field $\mathbf{S}_\textrm{E}$ minus the dimension of the L line/L surface which lies in $\mathbf{S}_\textrm{E}$, increases from one to two. 
The spin angular momentum of the field is zero when linearly polarised, meaning the direction of the normal to the field oscillations cannot be defined. Drawing a circuit around an L line, the spin vector rotates through $2\pi$ radians in a clockwise or anti-clockwise sense and defines the L line's topological index.\\
\indent The characteristics of scalar vortices and C lines and L lines are visualised in Fig.~\ref{fig:explain}.

\subsection{Indexing Point-like Singularities}
Polarisation singularities occur equally often among the general polarisation ellipses in $\mathbf{E}$ and $\mathbf{H}$ fields, and need not coincide with each other. 
Phase singularities, C lines and L lines are all indexed by looking at the circulation or rotation of a scalar or vector quantity around a loop enclosing the singularity of interest \cite{Berry2004}. 
All three of these singularities are threads in 3D fields, but the winding number concept can be generalised to higher dimensional singularities and calculated for point-like, 3D vector singularities via the topological degree. 
Instead of integrating a quantity associated with a line singularity around a 1D closed circuit, for isolated singular points in 3D, we should integrate an appropriate quantity over a closed surface enclosing the point singularity.
For a vector $\mathbf{V}(\mathbf{r}_S)$ on a surface $S$ ($\mathbf{r}_S\in S$) in 3D real space, for example, the topological degree of $\mathbf{r}_S\mapsto\mathbf{V}$ (the mapping from the real space surface $\mathbf{r}_S$ to $\mathbf{V}$) is a calculation of the integer number of times that every possible direction of $\mathbf{V}$ is realised (on a sphere) on all the points $\mathbf{r}_S$ on the surface $S$. 
As with other kinds of topological singularities in physical fields, the easiest realised topological degrees (winding numbers) are $\pm 1$.
Mathematically, a $0,\pm 1$ topological degree is the integral of the determinant of the dyadic $D(\mathbf{V})$ of $\mathbf{V}$ over $S$ divided by $A$, the area of $S$,
\begin{equation}\label{topdegree}
    \textrm{deg}(\mathbf{V})=\frac{1}{A}\int_S\textrm{det}(D(\mathbf{V}))dS.
\end{equation}
The dyadic $D(\mathbf{V})$, also called the Jacobian matrix of $\mathbf{V}$, contains the first order spatial derivatives of each component of $\mathbf{V}$. 
The sign of the determinant of $D(\mathbf{V})$ equals the product of the signs of its eigenvalues. 
For 3D vectors where $D(\mathbf{V})$ is a $3\times 3$ matrix, it is possible for drastically different behaviour of $\mathbf{V}$ to be hidden under the same topological degree. 
For example, if $\mathbf{V}(\mathbf{r=0})=\mathbf{0}$ (meaning the direction of $\mathbf{V}$ is singular at the origin) and we assume that a linear map from an origin-enclosing surface to $\mathbf{V}$ has a topological degree of $-1$, then $D(\mathbf{V})$ at $\mathbf{r=0}$ could have signed eigenvalues (in any order) of $++-$ or $---$. 
Physically, the origin could either be a saddle point or a sink for $\mathbf{V}$ with no distinction in topological degree because both $++-$ and $---$ eigenvalues multiply to a negative sign. 
Rather than calculating the topological degree, to try to classify the flux of energy and canonical momentum through a 3D optical field zero, we use the signs of the eigenvalues of their first order dyadics evaluated in the position of the field zero.\\
\indent We use the ideas discussed here to report our findings in the following sub-sections, beginning with the six possible ways that C lines and L lines can intersect in a 3D zero.

\subsection{Polarisation Singularities at a 3D Electric Field Zero}
\begin{figure}
    \centering
    \includegraphics[width=\textwidth]{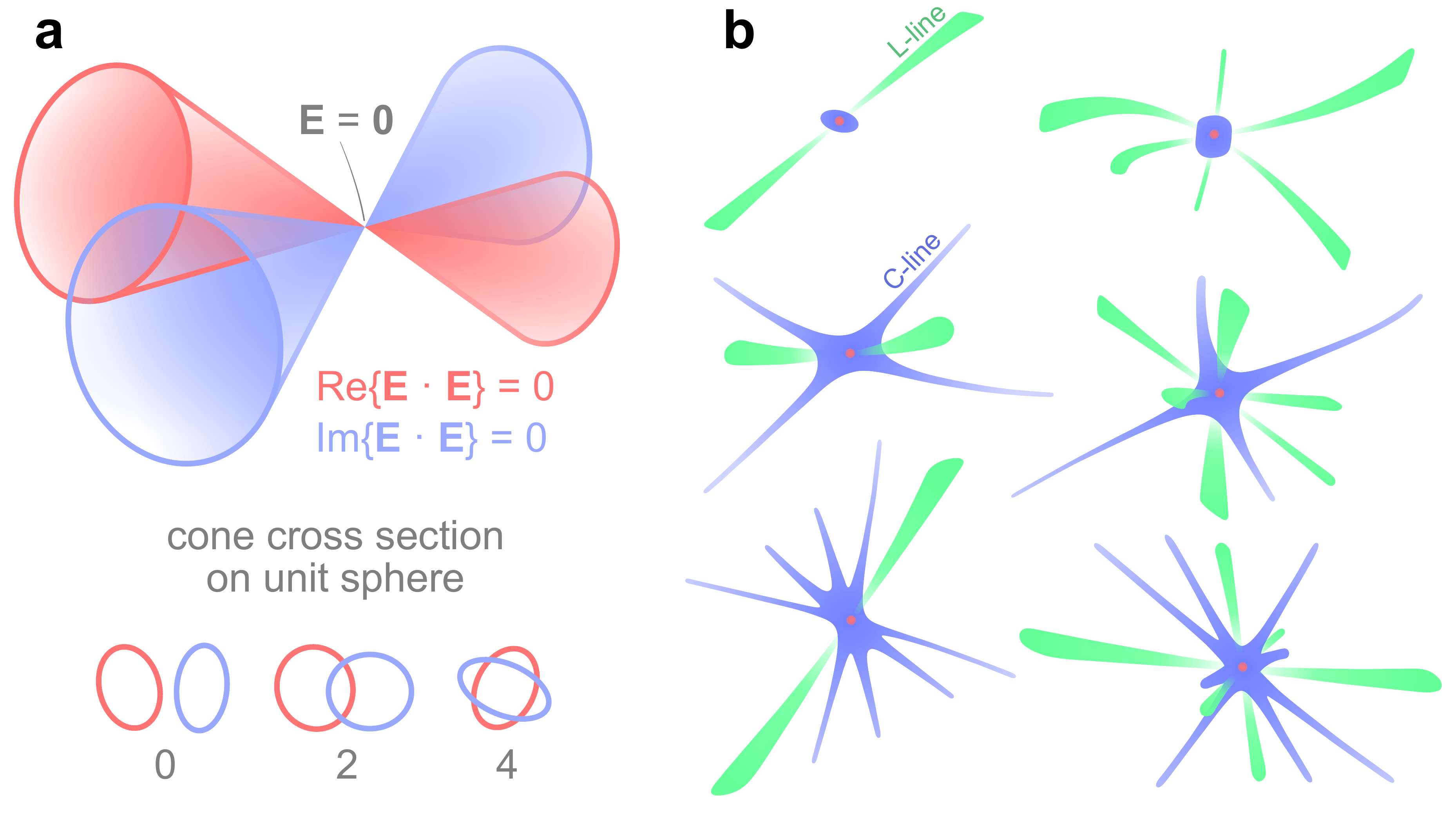}
    \caption{Electric polarisation singularities passing through a 3D electric field zero at a position $\mathbf{r}_0$. (a) Visualisation of why zero, two or four C lines must pass through $\mathbf{r}_0$. 
    In a first-order approximation, the surfaces $\textrm{Re}\{\mathbf{E\cdot E}\}=0$ (red) and $\textrm{Im}\{\mathbf{E\cdot E}\}=0$ (blue) are double cones, and where they intersect, C lines exist. 
    Two double cones intersect along two or four lines, or do not intersect at all, which is easy to see by considering the cones' cross sections on the unit sphere: ellipses which cross at zero, two or four points. 
    (b) Six different examples of electric field zeros created at a position $\mathbf{r}_0$ (red circle), one per unique combination of C lines and L lines meeting there. 
    The C lines are marked by blue regions where $\mathbf{E\cdot E}\approx0$ and the L lines by the green regions where $\textrm{Im}\{\mathbf{E^*\times E}\}\approx\mathbf{0}$. 
    Each field zero is created in analytical simulations by designing the polarisation of ten plane waves with random wavevectors, wavelength 500 nm, to interfere destructively at $\mathbf{r}_0$. 
    The plane waves have different polarisations and wavevectors for each example zero in (b).}
    \label{fig:types}
\end{figure}

We will focus on a 3D electric field zero in a position $\mathbf{r}_0$, that is $\mathbf{E}(\mathbf{r}_0)=\mathbf{0}$, and study the nearby strands of circular and linear electric polarisation. Identical arguments to those given here could be made for magnetic field zeros ($\mathbf{H}(\mathbf{r}_\textrm{0})=\mathbf{0}$) and magnetic polarisation singularities, or for simultaneous electric and magnetic field zeros ($\mathbf{E}(\mathbf{r}_\textrm{0})=\mathbf{H}(\mathbf{r}_\textrm{0})=\mathbf{0}$) and polarisation singularities of either $\mathbf{E}$ or $\mathbf{H}$. 
Any smooth function of $\mathbf{r}$ is nearly linear over small distances, which means all fundamental behaviour of the electric field in the immediate vicinity of the zero is captured by its Jacobian, $\mathbf{J}_\textrm{E}=D(\mathbf{E})$, a complex $3\times3$ matrix containing all first-order spatial derivatives of $E_x$, $E_y$ and $E_z$, evaluated at $\mathbf{r}_0$,

\begin{equation}\label{jacE}
    \mathbf{J}_\textrm{E}=D(\mathbf{E})=\begin{pmatrix}\frac{\partial E_x}{\partial x}&\frac{\partial E_x}{\partial y}&\frac{\partial E_x}{\partial z}\\
                                         \frac{\partial E_y}{\partial x}&\frac{\partial E_y}{\partial y}&\frac{\partial E_y}{\partial z}\\
                                         \frac{\partial E_z}{\partial x}&\frac{\partial E_z}{\partial y}&\frac{\partial E_z}{\partial z}\end{pmatrix}=(\grad\otimes\mathbf{E})^T.
\end{equation}
The Jacobian of the magnetic field at $\mathbf{r}_0$, $\mathbf{J}_\textrm{H}$, can be defined similarly. 
In free space, $\mathbf{J}_\textrm{E}$ and $\mathbf{J}_\textrm{H}$ are always traceless because $\mathbf{E}$ and $\mathbf{H}$ are divergence-free. 
Maxwell's equations also require that if $\mathbf{E}(\mathbf{r}_\textrm{0})=\mathbf{0}$, then $\mathbf{J}_\textrm{H}$ must be symmetric at $\mathbf{r}_\textrm{0}$ and vice versa for $\mathbf{H}(\mathbf{r}_\textrm{0})=\mathbf{0}$. 
We make a first-order approximation of the electric field vector near $\mathbf{r}_0$ with,
\begin{equation}\label{Eapprox}
    \mathbf{\tilde{E}}=\mathbf{J}_\textrm{E}\mathbf{v},
\end{equation}
where $\mathbf{v=r-}\mathbf{r}_0$.\\
\indent Nearby C lines emerge in our approximated field wherever $\mathbf{\tilde{E}\cdot \tilde{E}}=0$, which we may calculate using (\ref{Eapprox}) and separate into real and imaginary parts,
\begin{equation}\label{EdotEapprox}
    \begin{gathered}
    \mathbf{\tilde{E}\cdot \tilde{E}}=(\mathbf{J}_\textrm{E}\mathbf{v})\cdot(\mathbf{J}_\textrm{E}\mathbf{v})\\=\mathbf{v}^T\mathbf{M}\mathbf{v}+i\mathbf{v}^T\mathbf{N}\mathbf{v},
    \end{gathered}
\end{equation}
where $\mathbf{M}=\textrm{Re}\{\mathbf{J}_\textrm{E}^T\mathbf{J}_\textrm{E}\}$ and $\mathbf{N}=\textrm{Im}\{\mathbf{J}_\textrm{E}^T\mathbf{J}_\textrm{E}\}$. 
The two terms in equation (\ref{EdotEapprox}) are quadric surfaces connecting constant valued real and imaginary parts of $\mathbf{\tilde{E}\cdot \tilde{E}}$, and the real and imaginary surfaces described by setting (\ref{EdotEapprox}) equal to zero cross in real space where $\mathbf{\tilde{E}}$ is circularly polarised. 
The real $3\times3$ matrices $\mathbf{M}$ and $\mathbf{N}$ are symmetric and always have real eigenvalues. 
Normally, these eigenvalues have signs $++-$ or $--+$ (in any order) so that the surfaces $\mathbf{v}^T\mathbf{M}\mathbf{v}=0$ and $\mathbf{v}^T\mathbf{N}\mathbf{v}=0$ are both double cones, vertices touching at $\mathbf{v=0}$ as shown in Fig.~\ref{fig:types}(a). 
The cones have an elliptical cross section whose ellipticity is constant with distance from $\mathbf{v}=0$ in the linear approximation. 
Because two ellipses can intersect at either zero, two or four points (as shown in the lower part of Fig.~\ref{fig:types}(a)), there must be either zero, two or four C lines passing through the electric field zero. 
If one matrix, say $\mathbf{M}$, is positive or negative definite (all positive or all negative eigenvalues), $\textrm{Re}\{\mathbf{\tilde{E}\cdot \tilde{E}}\}$ will solely increase or decrease in all outward directions from $\mathbf{v=0}$. Then, the constant-valued surface $\mathbf{v}^T\mathbf{M}\mathbf{v}=C$ becomes an ellipsoid, and $\mathbf{v}^T\mathbf{M}\mathbf{v}=0$ is satisfied only at $\mathbf{v=0}$ so that no C lines pass through the 3D vortex.\\
\indent To reveal the number of L lines that extend through the 3D electric field zero, we must calculate the electric field spin, given by,
\begin{equation}\label{Espin}
    \mathbf{S}_\textrm{E}\propto\textrm{Im}\{\mathbf{E^*\times E}\}=2\textrm{Re}\{\mathbf{E}\}\times\textrm{Im}\{\mathbf{E}\}.
\end{equation}
When the electric field is linearly polarised ($\mathbf{S}_\textrm{E}=\mathbf{0}$), the real and imaginary parts of $\mathbf{E}$ must be (anti)parallel. 
Under the approximation (\ref{Eapprox}), this means,
\begin{equation}
\begin{gathered}
    \textrm{Re}\{\mathbf{J}_\textrm{E}\}\mathbf{v}=\lambda\textrm{Im}\{\mathbf{J}_\textrm{E}\}\mathbf{v},
\end{gathered}
\end{equation}
where $\lambda$ is a positive or negative scalar. 
The directions of the L lines crossing through $\mathbf{v=0}$ are given by the three eigenvectors of the matrix $\textrm{Im}\{\mathbf{J}_\textrm{E}\}^{-1}\textrm{Re}\{\mathbf{J}_\textrm{E}\}$. 
Since this matrix is real-valued, either all three of these eigenvectors are real, corresponding to three L lines, or only one of them is real and is accompanied by a conjugate pair of complex eigenvectors. 
In that case, just one L line passes through the 3D zero because $\mathbf{v}$ cannot point in a complex direction.\\
\indent Summarising, either zero, two or four C lines and either one or three L lines always meet at $\mathbf{r}_0$ in a 3D electric field zero $\mathbf{E}(\mathbf{r}_0)=\mathbf{0}$. 
An identical conclusion can be drawn for C lines and L lines of the magnetic field for the case of $\mathbf{H}(\mathbf{r}_0)=\mathbf{0}$. 
In Fig.~\ref{fig:types}(b), an example of each of the six possible C line/L line combinations through a 3D zero is presented, the zeros created in the interference of ten plane waves. 
Each zero is enforced by separate ensembles of ten plane waves with random wavevector directions that are deliberately polarised to destructively interfere at a single point.
\\
\subsection{Energy Flux Singularity}
\begin{figure}
    \centering
    \includegraphics[width=\textwidth]{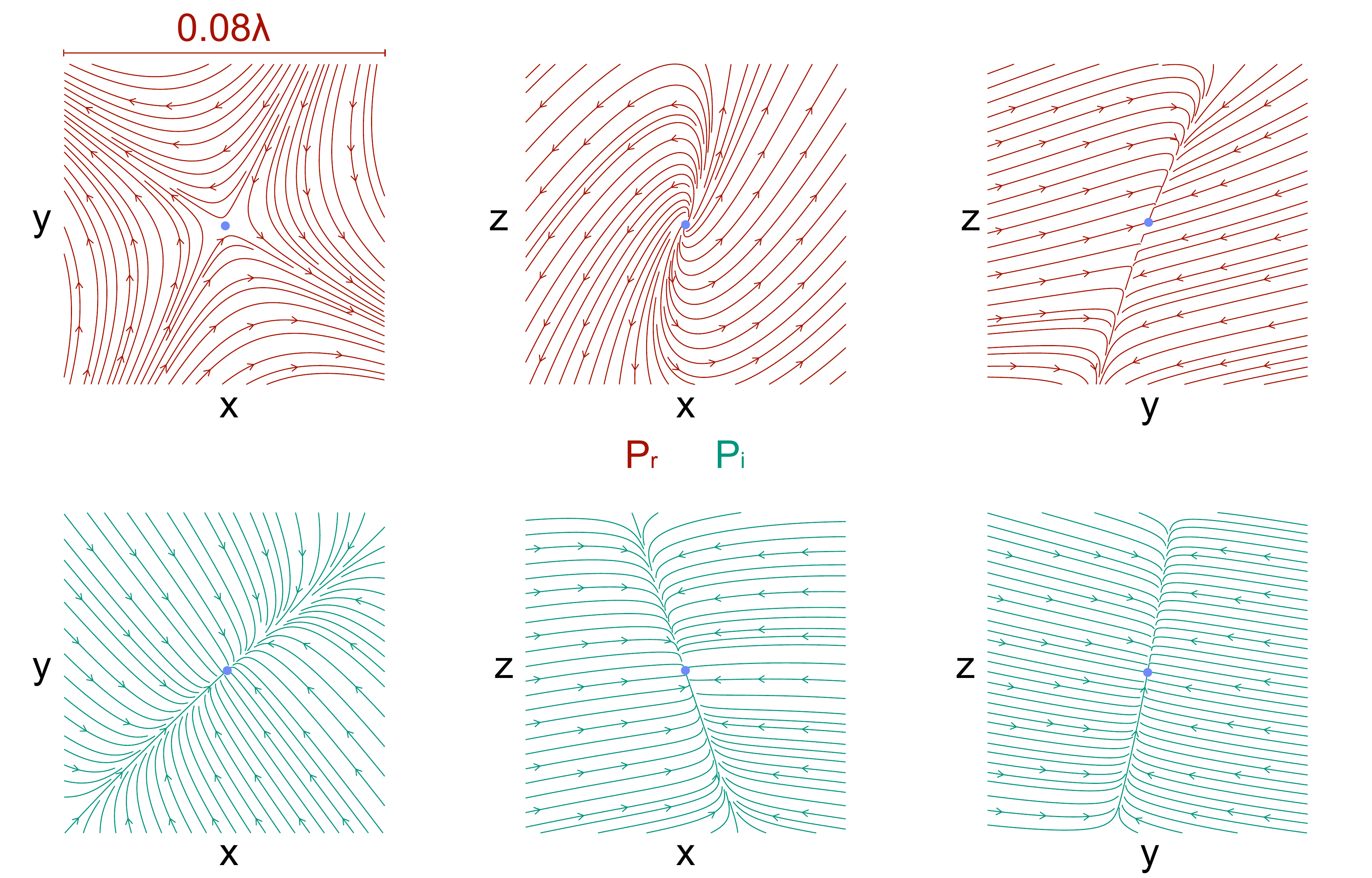}
    \caption{Flow of the real ($\mathbf{P}_\textrm{r}$, red) and imaginary ($\mathbf{P}_\textrm{i}$, teal) parts of the Poynting vector, $\frac{1}{2}\mathbf{E^*\times H}$, on the $xy$, $xz$ and $yz$ planes coinciding with an electric field zero at position $\mathbf{r}_0$ (blue circle). 
    The real Poynting vector is divergence-free, meaning a vector saddle point of $\mathbf{P}_\textrm{r}$ is set up at $\mathbf{r}_0$. 
    The imaginary Poynting vector is not necessarily divergence-free and can be organised in a sink at $\mathbf{r}_0$ when $\mathbf{E}(\mathbf{r}_0)=\mathbf{0}$ (a source is not possible unless the magnetic field is zero). 
    Results are generated by designing the polarisation of ten plane waves with random propagation directions to interfere completely at $\mathbf{r}_0$.}
    \label{fig:poynt}
\end{figure}

The flow of energy in a light field is described by the complex Poynting vector, $\frac{1}{2}\mathbf{E^*\times H}$. 
The real part of this vector (often itself called the `Poynting vector') corresponds to the time-averaged power transfer (sometimes known as active power) in the field, while reactive power (associated with oscillations in the transfer of power) is accounted for by the less-used imaginary part. 
We refer to these two real vectors as $\mathbf{P}_\textrm{r}$ and $\mathbf{P}_\textrm{i}$,
\begin{equation}\label{preal}
    \mathbf{P}_\textrm{r}=\frac{1}{2}\textrm{Re}\{\mathbf{E^*\times H}\}
\end{equation}
\begin{equation}
    \mathbf{P}_\textrm{i}=\frac{1}{2}\textrm{Im}\{\mathbf{E^*\times H}\}
\end{equation}
When either $\mathbf{E}$ or $\mathbf{H}$ is zero at a point $\mathbf{r}_0$, the complex Poynting vector vanishes, and its real and imaginary parts circulate in the space around the zero according to their first-order derivatives at $\mathbf{r}_0$. 
The real part $\mathbf{P}_\textrm{r}$ is divergence-less in free space where there is no absorption or energy generation, and must therefore be organised into a vector saddle point at $\mathbf{r}_0$. 
An example flow of active power around a 3D electric field zero created at $\mathbf{r}_0$ ($\mathbf{E}(\mathbf{r}_0)=\mathbf{0}$, $\mathbf{H}(\mathbf{r}_0)\neq\mathbf{0}$) is given in the top row of panels in Fig.~\ref{fig:poynt}, where $\mathbf{P}_\textrm{r}$ is plotted on the $xy$, $xz$, and $yz$ planes coinciding at $\mathbf{r}_0$. 
Although there is no net flow of active power in or out of the zero, $\mathbf{P}_\textrm{r}$ streamlines can be arranged in two topologically different ways depending on whether the signs of the eigenvalues of its first-order dyadic, $\textrm{Im}\{(\mathbf{J}_\textrm{E}^T-\mathbf{J}_\textrm{E})\mathbf{J}_\textrm{E}^*\}$ (written electrically without prefactors), are $++-$ or $+--$, corresponding to two possible topological orders of $-1$ or $+1$. 
One might notice that the imaginary Poynting vector $\mathbf{P}_\textrm{i}$, which is plotted on the same planes for the same free space electric field zero at $\mathbf{r}_0$ in the lower row of panels of Fig.~\ref{fig:poynt}, is not divergence-free---in fact, it is physically possible for a source, sink or saddle of $\mathbf{P}_\textrm{i}$ to exist there, depending on whether $\mathbf{E}$ or $\mathbf{H}$ is zero. 
To see why, we first note that using Maxwell's equations in free space (see supplemental information), the imaginary Poynting vector can be decomposed into a sum of two terms, one polarisation-independent and one polarisation-dependent, each containing electric and magnetic contributions,
\begin{equation}\label{pidecomp}
\begin{gathered}
    \mathbf{P}_\textrm{i}=-\frac{c^2}{2\omega}\epsilon_0\textrm{Re}\{(\mathbf{J}_\textrm{E}^T-\mathbf{J}_\textrm{E})\mathbf{E}^*\}\\
    =\frac{c^2}{2\omega}\mu_0\textrm{Re}\{(\mathbf{J}_\textrm{H}^T-\mathbf{J}_\textrm{H})\mathbf{H}^*\}\\
    =\frac{c^2}{4\omega}\left[-\frac{1}{2}\epsilon_0\grad(\mathbf{E^*\cdot E})+\frac{1}{2}\mu_0\grad(\mathbf{H^*\cdot H})\right]+\frac{c^2}{4\omega}\textrm{Re}\{\epsilon_0\mathbf{J}_\textrm{E}\mathbf{E}^*-\mu_0\mathbf{J}_\textrm{H}\mathbf{H}^*\}.
    \end{gathered}
\end{equation}
The first term in Eq.~(\ref{pidecomp}) represents the difference in gradient of the electric and magnetic energy density of the light field, while polarisation-dependent behaviour of $\mathbf{P}_\textrm{i}$ derives from the second term since $\mathbf{J}_\textrm{E}\mathbf{E}^*$ and $\mathbf{J}_\textrm{H}\mathbf{H}^*$ contain inter-component multiplication. 
In certain cases such as a uniformly polarised standing wave, the second term is zero and the gradient of the difference in electric and magnetic energy density determines the direction of reactive power flow. 
Because $\mathbf{E^*\cdot E}=|\mathbf{E}|^2$ is a positive real quantity, a 3D zero in $\mathbf{E}$ is a source for the vector $\grad(\mathbf{E^*\cdot E})$ (and likewise for $\mathbf{H}$). 
Depending on how the polarisation-independent and polarisation-dependent terms combine in Eq.~(\ref{pidecomp}), the imaginary Poynting vector could have non-zero divergence at $\mathbf{r}_0$. 
Note that there is a difference in sign between the electric and magnetic terms in Eq.~(\ref{pidecomp}), meaning $\mathbf{P}_\textrm{i}$ behaves differently for $\mathbf{E}(\mathbf{r}_0)=\mathbf{0}$, $\mathbf{H}(\mathbf{r}_0)\neq\mathbf{0}$ and $\mathbf{H}(\mathbf{r}_0)=\mathbf{0}$, $\mathbf{E}(\mathbf{r}_0)\neq\mathbf{0}$ and $\mathbf{E}(\mathbf{r}_0)=\mathbf{H}(\mathbf{r}_0)=\mathbf{0}$ 3D zeros. 
To understand the flow of $\mathbf{P}_\textrm{i}$ through an optical field zero, we assume a non-dual electric field zero ($\mathbf{E}(\mathbf{r}_0)=\mathbf{0}$ and $\mathbf{H}(\mathbf{r_0})\neq\mathbf{0}$) and make a first-order approximation of $\mathbf{P}_\textrm{i}$, this time referring to the relevant linear transformation matrix as the first-order dyadic of the imaginary Poynting vector, $D(\mathbf{P}_\textrm{i})$, which is defined identically to $\mathbf{J}_\textrm{E}$ in Eq.~(\ref{jacE}) with $\mathbf{P}_\textrm{i}$ and its components in place of $\mathbf{E}$. 
Our approximate imaginary Poynting vector is,
\begin{equation}\label{Piapprox}
    \mathbf{\tilde{P}}_\textrm{i}=D(\mathbf{P}_\textrm{i})\mathbf{v}
\end{equation}
where $\mathbf{v=r-r}_0$. 
The dyadic $D(\mathbf{P}_\textrm{i})=(\grad\otimes\mathbf{P}_\textrm{i})^T$ evaluated at $\mathbf{r}_0$ is, using the electric representation of $\mathbf{P}_\textrm{i}$ in Eq.~(\ref{pidecomp}) (top line),
\begin{equation}\label{DPi}
    D(\mathbf{P}_\textrm{i})=-\frac{c^2}{2\omega}\epsilon_0\textrm{Re}\{(\mathbf{J}_\textrm{E}^T-\mathbf{J}_\textrm{E})\mathbf{J}_\textrm{E}^*\}.
\end{equation}
There are no second order derivatives of $\mathbf{E}$ in Eq.~(\ref{DPi}) because $\mathbf{E}(\mathbf{r}_0)=\mathbf{0}$. 
Surprisingly, $D(\mathbf{P}_\textrm{i})$ cannot have three positive eigenvalues, as justified in the supplemental information. 
The result is that at a 3D electric field zero, $\mathbf{P}_\textrm{i}$ is organised into one of two types of saddle with topological degree $1$ or $-1$, or a sink with topological degree $-1$, never a source. 
When $\mathbf{H}(\mathbf{r}_0)=\mathbf{0}$ and $\mathbf{E}(\mathbf{r_0})\neq\mathbf{0}$, the opposite is true because of the dual-asymmetry of the imaginary Poynting vector: $\mathbf{P}_\textrm{i}$ can form a saddle or source at $\mathbf{r}_0$ but not a sink.
\\
\subsection{Orbital Current Singularity}
When divided by $c^2$, the real Poynting vector Eq.~(\ref{preal}) turns into a momentum density, the kinetic momentum density, which, using Maxwell's equations for time-harmonic fields, can be split in to a well-known sum of separate orbit and spin contributions \cite{Berry2009,Bliokh2017}. 
For instance, by substituting (with prefactors) the curl of $\mathbf{E}$ for $\mathbf{H}$, the kinetic momentum density can be written as,

\begin{equation}\label{kinmom}
\begin{gathered}
    \mathbf{\Pi}=\frac{1}{2c^2}\textrm{Re}\{\mathbf{E^*\times H}\}\\=\frac{1}{2\omega}\epsilon_0\textrm{Im}\{\mathbf{E^*}\cdot(\grad)\mathbf{E}\}+\frac{1}{2\omega}\epsilon_0\grad\times\frac{1}{2}\textrm{Im}\{\mathbf{E^*\times E}\},
\end{gathered}
\end{equation}
where $\mathbf{A}\cdot(\grad)\mathbf{B}=A_x\grad B_x+A_y\grad B_y+A_z\grad B_z=\mathbf{J}_\textrm{B}^T\mathbf{A}$, with $\mathbf{J}_\textrm{B}$ being the Jacobian of $\mathbf{B}$ defined identically to Eq.~(\ref{jacE}) (the decomposition is explained in more detail in the supplemental information). 
The first decomposed term is $\mathbf{p}_\textrm{E}^\textrm{o}$, the orbital contribution to the kinetic momentum density, called the canonical momentum density, imparted by the electric field only,
\begin{equation}\label{eleccanmom}
    \mathbf{p}^\textrm{o}_\textrm{E}=\frac{1}{2\omega}\epsilon_0\textrm{Im}\{\mathbf{E^*}\cdot(\grad)\mathbf{E}\}=\frac{1}{2\omega}\epsilon_0\textrm{Im}\{\mathbf{J}_\textrm{E}^T\mathbf{E}^*\}.
\end{equation}
Eq.~(\ref{kinmom}) can also be written purely in terms of $\mathbf{H}$ and by averaging these equivalent representations of $\mathbf{\Pi}$, the dual-symmetric canonical momentum density is obtained,
\begin{equation}\label{canmom}
    \mathbf{p}^\textrm{o}=\frac{1}{4\omega}\textrm{Im}\{\epsilon_0\mathbf{E^*}\cdot(\grad)\mathbf{E}+\mu_0\mathbf{H^*}\cdot(\grad)\mathbf{H}\}.
\end{equation}
This momentum density definition contains both the electric and magnetic field's influence, and produces the total orbital angular momentum of the field within a volume when $\mathbf{r}\times\mathbf{p}^\textrm{o}$ is integrated. 
Naturally, the electric and magnetic contributions to (\ref{canmom}) become zero whenever $\mathbf{E=0}$ and $\mathbf{H=0}$. 
This means that, in a 3D electric field zero positioned at $\mathbf{r}_0$, the direction of the electric contribution $\mathbf{p}_\textrm{E}^\textrm{o}$ is undefined and should circulate around $\mathbf{r}_0$ in some fashion. 
Of course, while the total canonical momentum density at $\mathbf{r}_0$ is not zero when only $\mathbf{E=0}$, we could draw the same conclusions we make here for Eq.~(\ref{canmom}) rather than Eq.~(\ref{eleccanmom}) near a dual 3D vortex ($\mathbf{E}(\mathbf{r}_0)=\mathbf{H}(\mathbf{r}_0)=\mathbf{0}$).
Note that by normalising $\mathbf{E}$, the argument to $\textrm{Im}\{\}$ in Eq.~(\ref{eleccanmom}) defines the local electric wavevector \cite{Berry2001},
\begin{equation}\label{kloc}
    \mathbf{k}_\textrm{loc}^\textrm{e}=-i\mathbf{e^*}\cdot(\grad)\mathbf{e},
\end{equation}
\begin{figure}
    \centering
    \includegraphics[width=\textwidth]{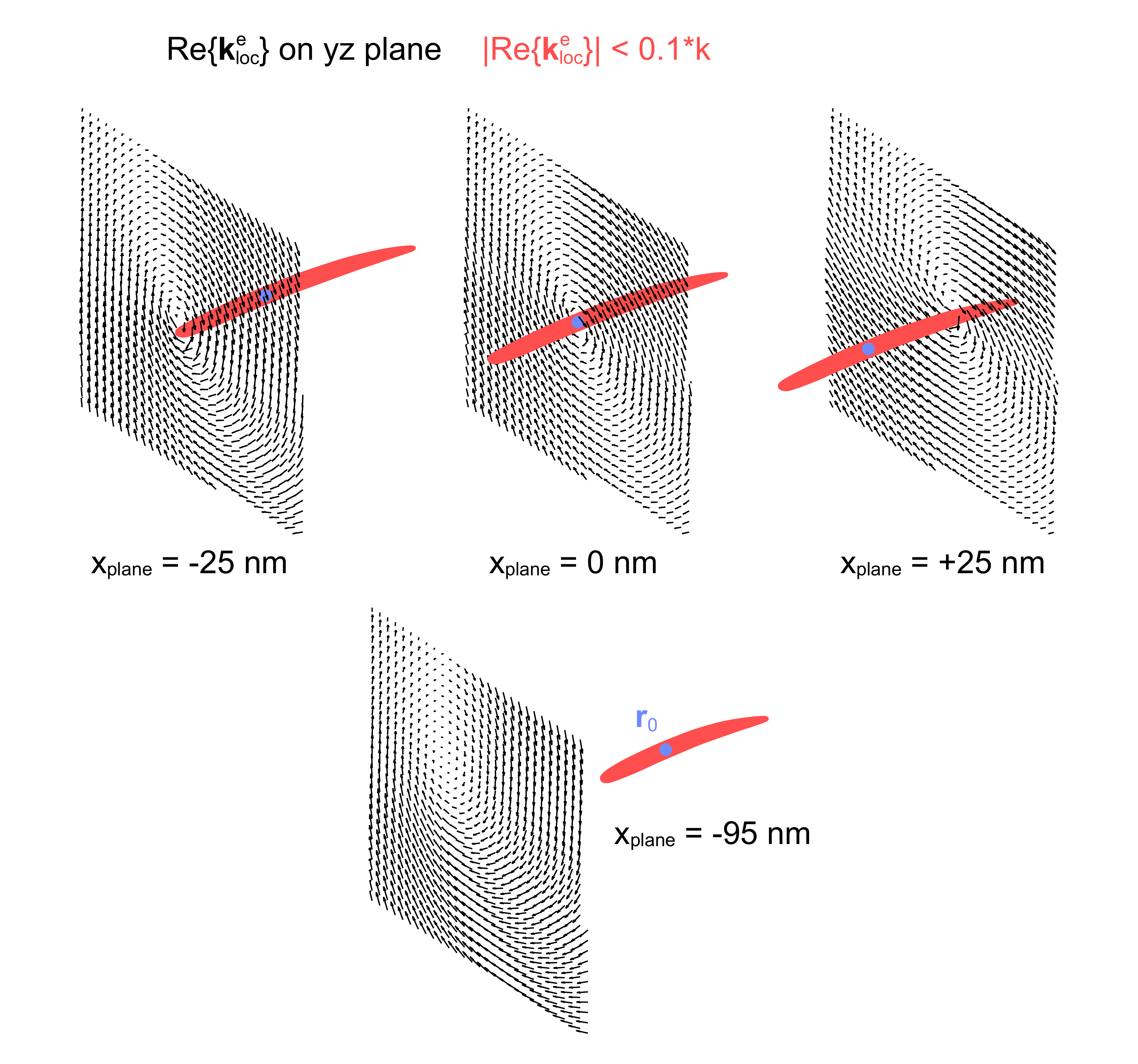}
    \caption{Vortex pseudo-line (red) of the real electric local wavevector, $\textrm{Re}\{\mathbf{k}_\textrm{loc}^\textrm{e}\}$, passing though an electric field zero at position $\mathbf{r}_0$ (blue circle), that is $\mathbf{E}(\mathbf{r}_0)=\mathbf{0}$ and $\mathbf{H}(\mathbf{r}_0)\neq\mathbf{0}$. 
    The red line indicates regions of space where $|\textrm{Re}\{\mathbf{k}_\textrm{loc}^\textrm{e}\}|<0.1k$, where $k=\frac{2\pi}{\lambda}$ and $\lambda=500$ nm. 
    The line is roughly oriented along the $x$ axis and the electric local wavevector is plotted on four different $yz$ planes. 
    The three planes which coincide with the line are $-25$ nm, 0 nm, $+25$ nm in the $x$ direction away from $\mathbf{r}_0$, showing clear vortex-like circulation of momentum around the axis of the red line. 
    On the fourth plane $-95$ nm away from $\mathbf{r}_0$, the vortex-like circulation of $\textrm{Re}\{\mathbf{k}_\textrm{loc}^\textrm{e}\}$ has lost some definition, highlighting that $\textrm{Re}\{\mathbf{k}_\textrm{loc}^\textrm{e}\}$ is not exactly zero along a line, and only appears line-like near to the $\mathbf{E}$ field zero (the only location where $\textrm{Re}\{\mathbf{k}_\textrm{loc}^\textrm{e}\}$ is exactly zero is at $\mathbf{r}_0$, because $\textrm{Re}\{\mathbf{k}_\textrm{loc}^\textrm{e}\}$ vanishes at points, not along lines).  
    Results are generated from interference of ten plane waves with random wavevectors, wavelength $\lambda=500$ nm, deliberately polarised to create a 3D electric field zero at $\mathbf{r}_0$.}
    \label{fig:k}
\end{figure}
where $\mathbf{e}=\frac{\mathbf{E}}{\sqrt{\mathbf{E^*\cdot E}}}$. 
The real part of $\mathbf{k}_\textrm{loc}^\textrm{e}$ is the local phase gradient of the electric field, while $\textrm{Im}\{\mathbf{k}_\textrm{loc}^\textrm{e}\}$ points in the direction of decreasing electric field intensity. 
A three-dimensional, real vector, $\textrm{Re}\{\mathbf{k}_\textrm{loc}^\textrm{E}\}$ (and therefore canonical momentum density) can vanish at localised points in space with non-zero electric field, where a saddle-like circulation of $\textrm{Re}\{\mathbf{k}_\textrm{loc}^\textrm{E}\}$ surrounds \cite{Berry2019}, similar to the top row of panels in Fig.~\ref{fig:poynt}. 
But when the electric field vanishes and the direction of $\textrm{Re}\{\mathbf{k}_\textrm{loc}^\textrm{E}\}$ is automatically undefined, a different behaviour emerges.\\
\indent To understand why, we once again make a first-order approximation, this time of $\mathbf{p}_\textrm{E}^\textrm{o}$, capturing the electric canonical momentum very near to a 3D electric field zero at $\mathbf{r}_0$ in its dyadic $D(\mathbf{p}^\textrm{o}_\textrm{E})$,
\begin{equation}\label{poeapprox}
    \mathbf{\tilde{p}}^\textrm{o}_\textrm{E}=D(\mathbf{p}^\textrm{o}_\textrm{E})\mathbf{v},
\end{equation}
where $\mathbf{v=r-r}_0$. The dyadic $D(\mathbf{p}^\textrm{o}_\textrm{E})=(\grad\otimes\mathbf{p}^\textrm{o}_\textrm{E})^T$ at a general point in space is given by,
\begin{equation}\label{dyadic}
    D(\mathbf{p}^\textrm{o}_\textrm{E})=\frac{1}{4\omega}\epsilon_0\textrm{Im}\{\mathbf{J}_\textrm{E}^T\mathbf{J}_\textrm{E}^*\}+\frac{1}{4\omega}\epsilon_0\textrm{Im}\{E_x^*\textrm{Hess}(E_x)+E_y^*\textrm{Hess}(E_y)+E_z^*\textrm{Hess}(E_z)\},
\end{equation}
where $\textrm{Hess}(A)$ is the Hessian matrix of the scalar field $A$,
\begin{equation}
    \textrm{Hess}(A)=\begin{pmatrix}\frac{\partial^2A}{\partial x^2} & \frac{\partial^2A}{\partial x\partial y} & \frac{\partial^2A}{\partial x\partial z}\\
                          \frac{\partial^2A}{\partial y\partial x} & \frac{\partial^2A}{\partial y^2} & \frac{\partial^2A}{\partial y\partial z}\\
                          \frac{\partial^2A}{\partial z\partial x} & \frac{\partial^2A}{\partial z\partial y} & \frac{\partial^2A}{\partial z^2}
    \end{pmatrix}.
\end{equation}
As $\mathbf{E}$ approaches zero, the trace-less matrix $D(\mathbf{p}^\textrm{o}_\textrm{E})$ is dominated by the first term in Eq.~(\ref{dyadic}) and if evaluated at a location $\mathbf{r}_0$ where $\mathbf{E}(\mathbf{r}_0)=\mathbf{0}$, the linear approximation of $\mathbf{p}^\textrm{o}_\textrm{E}$ responds only to the properties of the matrix in the first term of Eq.~(\ref{dyadic}), $\textrm{Im}\{\mathbf{J}_\textrm{E}^T\mathbf{J}_\textrm{E}^*\}$. 
This is an anti-symmetric matrix which always has one zero and two purely imaginary eigenvalues, meaning that in the direction of the one real eigenvector of $D(\mathbf{p}^\textrm{o}_\textrm{E})$ at $\mathbf{r}_0$, the approximated electric canonical momentum does not increase at all, producing a zero-momentum line. 
The imaginary eigenvalues of $D(\mathbf{p}^\textrm{o}_\textrm{E})$ twists $\mathbf{p}^\textrm{o}_\textrm{E}$ into a surrounding vortex-like structure. 
This special type of vector field singularity is called a circulation. 
Fundamentally, the canonical momentum should only be zero at confined points in general 3D fields, so this apparent vortex line is only preserved locally to the electric field zero at $\mathbf{r}_0$, dissolving with distance as higher-order derivatives of $\mathbf{p}^\textrm{o}_\textrm{E}$ become significant (it is, in fact, just a very elongated null point of $\mathbf{p}^\textrm{o}_\textrm{E}$). 
The direction of the vortex pseudo-line in the vicinity of the electric field zero is also given by the curl of the orbital current,
\begin{equation}
    \mathbf{D}=\grad\times\mathbf{p}^\textrm{o}_\textrm{E}\propto\textrm{Re}\{\grad E_x\}\times\textrm{Im}\{\grad E_x\}+\textrm{Re}\{\grad E_y\}\times\textrm{Im}\{\grad E_y\}+\textrm{Re}\{\grad E_z\}\times\textrm{Im}\{\grad E_z\}.
\end{equation}
We visualise this feature in Fig.~\ref{fig:k}, where a 3D electric field zero is created at a point $\mathbf{r}_0$ by deliberately polarising ten plane waves, each with random wavevectors, to destructively interfere at $\mathbf{r}_0$. 
The real part of the electric local wavevector, $\textrm{Re}\{\mathbf{k}_\textrm{loc}^\textrm{e}\}$, the real part of Eq.~(\ref{kloc}), is calculated and the region of space where $|\textrm{Re}\{\mathbf{k}_\textrm{loc}^\textrm{e}\}|<0.1k$ ($k=\frac{2\pi}{\lambda}$) is revealed by a red line approximately $0.1\lambda$ in length. 
The electric local wavevector is proportional to $\mathbf{p}^\textrm{o}_\textrm{E}$ and shows the direction of canonical momentum carried by the electric field. 
This red line is not continuous; $\textrm{Re}\{\mathbf{k}_\textrm{loc}^\textrm{e}\}$ actually vanishes only at $\mathbf{r}_0$ but it increases in magnitude so slowly in a certain direction (the direction of the real eigenvector of $\textrm{Im}\{\mathbf{J}_\textrm{E}^T\mathbf{J}_\textrm{E}^*\}$) that a line-like structure of $|\textrm{Re}\{\mathbf{k}_\textrm{loc}^\textrm{e}\}|\approx0$ exists very near to $\mathbf{r}_0$, stirring the electric canonical momentum into a local vortex. 
This is shown by the four $yz$ planes on which $\textrm{Re}\{\mathbf{k}_\textrm{loc}^\textrm{e}\}$ is plotted in Fig.~\ref{fig:k}. 
The real part of the electric local wavevector forms a swirl around the red line, a swirl losing definition if the plotting plane is too far from $\mathbf{r}_0$. 
This remarkable structure always appears when all three electric field components are zero together at a point.
\\
\subsection{Spin Current}
In the decomposition of the kinetic momentum density, Eq.~(\ref{kinmom}), the second term is called the spin momentum. 
It is proportional (and should not be confused with) the curl of the spin angular momentum of the electric, magnetic or electromagnetic field depending on the representation. 
Like before, we will focus on the electric representation of the decomposed kinetic momentum density, referring to the electric spin momentum with $\mathbf{p}^\textrm{s}_\textrm{E}$,
\begin{equation}\label{spinmom}
    \mathbf{p}^\textrm{s}_\textrm{E}=\frac{1}{2\omega}\epsilon_0\grad\times\frac{1}{2}\textrm{Im}\{\mathbf{E^*\times E}\}=-\frac{1}{2\omega}\epsilon_0\textrm{Im}\{\mathbf{J}_\textrm{E}\mathbf{E}^*\}.
\end{equation}
The electric spin momentum is a divergence-free vector whose dyadic $D(\mathbf{p}^\textrm{s}_\textrm{E})$ has three non-zero eigenvalues when evaluated in the position of an electric field zero, organising $\mathbf{p}^\textrm{s}_\textrm{E}$ into one of two types of 3D vector saddle point, just like the real Poynting vector in Fig.~\ref{fig:poynt}. 
Expressing, in Eq.~(\ref{spinmom}), the electric spin momentum with the electric field Jacobian reveals that only a difference in sign and orientation of $\mathbf{J}_\textrm{E}$ separates $\mathbf{p}^\textrm{s}_\textrm{E}$ from the electric canonical momentum $\mathbf{p}^\textrm{o}_\textrm{E}$, given by Eq.~(\ref{eleccanmom}). 
This means that, in a dual electric-magnetic zero, $\mathbf{E}(\mathbf{r}_0)=\mathbf{H}(\mathbf{r}_0)=\mathbf{0}$, where $\mathbf{J}_\textrm{E}$ is symmetric from Maxwell's equations, the spin and canonical momentum dyadics are equal and opposite, $D(\mathbf{p}^\textrm{s}_\textrm{E})=-D(\mathbf{p}^\textrm{o}_\textrm{E})$ (this also means that the dyadic of the real Poynting vector is zero). 
In a first-order approximation of both $\mathbf{p}^\textrm{s}_\textrm{E}$ and $\mathbf{p}^\textrm{o}_\textrm{E}$ near $\mathbf{r}_0$ in this case, a zero-line exists in exactly the same place for both vectors, and around it, $\mathbf{p}^\textrm{s}_\textrm{E}$ and $\mathbf{p}^\textrm{o}_\textrm{E}$ have vortex-like circulation with opposite handedness to each other.
\\
\subsection{Spin Angular Momentum}
The dual spin angular momentum, created by the rotation of the electric and magnetic field vectors, is given by \cite{Bekshaev2015},
\begin{equation}\label{fullspin}
    \mathbf{S}=\frac{1}{4\omega}\textrm{Im}\{\epsilon_0\mathbf{E^*\times E}+\mu_0\mathbf{H^*\times H}\}.
\end{equation}
The electric and magnetic parts individually describe the ellipticity of the electric and magnetic polarisation ellipses, pointing in the perpendicular direction to the ellipse plane. Once more for simplicity, we will focus on the singularity in the electric field spin angular momentum, $\mathbf{S}_\textrm{E}=\frac{1}{4\omega}\textrm{Im}\{\epsilon_0\mathbf{E^*\times E}\}$, left in a 3D electric field zero positioned at $\mathbf{r}_0$. 
The total spin angular momentum, Eq.~(\ref{fullspin}), is not zero if only $\mathbf{E}(\mathbf{r}_0)=\mathbf{0}$, but we could draw similar conclusions for $\mathbf{S}$ as we do here for $\mathbf{S}_\textrm{E}$ when the electric and magnetic fields are simultaneously zero at $\mathbf{r}_0$.

Decomposing $\mathbf{S}_\textrm{E}$ using Maxwell's equations, we can write its first-order dyadic at $\mathbf{r}_0$ in terms of the light field Jacobian matrices (see supplemental material),
\begin{equation}\label{EspinDyad}
    D(\mathbf{S}_\textrm{E})=\frac{1}{4\omega^2}\epsilon_0\textrm{Re}\{(\mathbf{J}_\textrm{H}^T-\mathbf{J}_\textrm{H})\mathbf{J}^*_\textrm{E}\}.
\end{equation}
We note that Eq.~(\ref{EspinDyad}), describing the spatial derivatives of the electric field spin only, depends on the magnetic field Jacobian matrix $\mathbf{J}_\textrm{H}$, which is automatically symmetric whenever $\mathbf{E=0}$ from Maxwell's equations. 
The consequence is that $\mathbf{J}_\textrm{H}^T-\mathbf{J}_\textrm{H}=\mathbf{0}$ and all elements of $D(\mathbf{S}_\textrm{E})$ at $\mathbf{r}_0$ are zero when $\mathbf{E}(\mathbf{r}_0)=\mathbf{0}$. 
Higher-order derivatives of $\mathbf{S}_\textrm{E}$ (Hessian matrices for each component) need to be calculated to fully understand the flux of the electric spin angular momentum in the neighbourhood of a 3D zero in $\mathbf{E}$.\\

\subsection{Summary Table}
Here, in Table 1, we summarise the seven dyadics which classify the number of crossing C lines and L lines, the flux of the real and imaginary parts of the Poynting vector, the spin current, and the orientation of the canonical momentum vortex pseudo-line existing at a 3D electric field zero, $\mathbf{E}(\mathbf{r}_0)=\mathbf{0}$ while $\mathbf{H}(\mathbf{r}_0)\neq\mathbf{0}$. 
To characterise a magnetic field zero, the matrices can be written magnetically by substituting $\mathbf{J}_\textrm{E}$ for $\mathbf{J}_\textrm{H}$ (and changing the `$-$' sign in front of matrix 7 to a `$+$'), in which case matrices 1, 2, and 3 characterise magnetic polarisation singularities, and matrix 4 and 5 the magnetic local wavevector and spin current respectively. 
In the case of a dual 3D zero, $\mathbf{E}(\mathbf{r}_0)=\mathbf{H}(\mathbf{r}_0)=\mathbf{0}$, matrices 6 and 7 are zero because both $\mathbf{J}_\textrm{E}$ and $\mathbf{J}_\textrm{H}$ are symmetric. 
\begin{table}
\centering
\resizebox{\textwidth}{!}{
\begin{tabular}{c c c}
    & Matrix at $\mathbf{r}_0$ & Characteristic\\
    \hline\\
    & $\mathbf{J}_\textrm{E}$ & 3D complex Jacobian of the electric field at $\mathbf{r}_0$ (Eq.~(\ref{jacE}))\\\\
    1. & $\textrm{Im}\{\mathbf{J}_\textrm{E}\}^{-1}\textrm{Re}\{\mathbf{J}_\textrm{E}\}$ & Number of real eigenvalues is the number of L lines passing through $\mathbf{r}_0$\\\\
    2. & $\textrm{Re}\{\mathbf{J}_\textrm{E}^T\mathbf{J}_\textrm{E}\}$ & \makecell{Eigenvectors are the principle axes of the double cone $\textrm{Re}\{\mathbf{E\cdot E}\}=0$.\\ Number of intersections of this double cone with that of matrix 3 are the number of C lines.} \\\\
    3. & $\textrm{Im}\{\mathbf{J}_\textrm{E}^T\mathbf{J}_\textrm{E}\}$ & \makecell{Eigenvectors are the principle axes of the double cone $\textrm{Im}\{\mathbf{E\cdot E}\}=0$.\\ Number of intersections of this double cone with that of matrix 2 are the number of C lines.}\\\\
    4. & $\textrm{Im}\{\mathbf{J}_\textrm{E}^T\mathbf{J}_\textrm{E}^*\}$ & \makecell{Direction of real eigenvector (there is only one) is the axis of the electric local wavevector vortex.\\ Imaginary eigenvectors give the handedness of momentum circulation.}\\\\
    5. & $-\textrm{Im}\{\mathbf{J}_\textrm{E}\mathbf{J}_\textrm{E}^*\}$ & \makecell{Proportional to first-order dyadic of spin current (Eq.~(\ref{spinmom})).\\ Eigenvalue signs give the type of minimum at $\mathbf{r}_0$}\\\\
    6. & $\textrm{Im}\{(\mathbf{J}_\textrm{E}^T-\mathbf{J}_\textrm{E})\mathbf{J}_\textrm{E}^*\}$ & \makecell{Proportional to first-order dyadic of real Poynting vector (active power flow).\\ Eigenvalue signs give the type of minimum at $\mathbf{r}_0$}\\\\
    7. & $-\textrm{Re}\{(\mathbf{J}_\textrm{E}^T-\mathbf{J}_\textrm{E})\mathbf{J}_\textrm{E}^*\}$ & \makecell{Proportional to first-order dyadic of imaginary Poynting vector (reactive power flow).\\Eigenvalue signs give the type of minimum at $\mathbf{r}_0$}\\
    \hline
\end{tabular}}
\caption{Summary of the seven dyadics (numbered) which classify the vector field singularities organised by a 3D electric field zero.}
\end{table}

\section{Discussion}
Three-dimensional optical field zeros are co-dimension 6 entities which, unlike axial zeros in beams, are completely localised, the optical field growing brighter in all outward directions. 
Although they rarely occur naturally in light (requiring three additional parameters beyond spatial $x,y,z$ due to their codimension), 3D zeros can be deliberately created in plane wave interference or in the near fields of light-scattering matter \cite{Vernon2022} to reveal the unusual features they imprint in the light field's energy, wavevector and polarisation structures. 
Both with mathematical argument and by creating field zeros in plane wave interference, we showed that whenever the electric or magnetic field is zero at a point $\mathbf{r}_0$, then some combination of zero, two or four C lines, lines of pure circular polarisation, and one or three L lines, lines of pure linear polarisation of the field in question, intersect at $\mathbf{r}_0$. 
Likewise, an imprint is made at $\mathbf{r}_0$ in the surrounding flux of the parts of the complex Poynting vector $\frac{1}{2}\mathbf{E^*\times H}$, the local wavevector, the spin momentum and spin angular momentum, each organised in a vector source, sink or saddle point. 
The signs of the eigenvalues of the first-order dyadics of each quantity at $\mathbf{r}_0$ reveal this. 
Of particular interest is the canonical momentum: while typically vanishing at confined points in space, a zero in $\mathbf{E}$ or $\mathbf{H}$ at $\mathbf{r}_0$ twists the canonical momentum imparted by that null-containing field into a sub-wavelength, vortex-like structure around an axis with an easily calculated direction. 
We say it is a sub-wavelength object because, although it resembles the twisted vortex structures of well-known doughnut beams, it is not preserved with increasing distance from $\mathbf{r}_0$. 
In the combination of the way energy flows through $\mathbf{r}_0$ and the number of intersecting polarisation singularities, any 3D field zero inscribes one of a discrete number of topologically unique signatures in the electromagnetic field. 
We identify seven dyadics whose spectra could classify all physically possible imprints of 3D optical field zeros.\\
\indent It is tempting to speculate that a surface enclosing an electric or magnetic field point zero might, in addition to the quantities already identified, possess a nonzero topological Chern number due to a nontrivial geometric phase 2-form (Berry curvature) resulting from the neighbouring polarisation pattern.
The appropriate expression for the geometric phase 2-form is the curl of the local wavevector Eq. (\ref{kloc}), 
\begin{equation}\label{berrycurv}
   \mathbf{V} = \grad \times \mathbf{k}_\textrm{loc}^\textrm{e}
\end{equation}
Near an electric field zero, $\mathbf{V}$ is anti-symmetric; integrating over a small sphere centred on the field zero gives zero.
We showed that in its neighbourhood, a 3D zero in $\mathbf{E}$ constructs a local wavevector vortex with an identifiable axis along which $|\mathbf{V}|$ is very large.
It is interesting that even when the complete vector characteristics of light are considered, a linear momentum vortex line still persists when all three field components are zero at a confined point.
This vector field vortex is an analogue to a phase vortex in a complex scalar field, with a key difference being that the vector field vortex line is not continuous.
Although the electromagnetic zero has some topological effects as we described in this paper, it is not so strong as to endow a surface around it with a nonzero Chern number.\\
\indent We have shown that, despite being unstable to perturbation, 3D zeros of the electric and electromagnetic field have topological properties generalising those of scalar vortices and polarisation singularities.
Further studies might indicate how these properties behave under perturbation.
We hope that by highlighting the unusual properties of 3D field zeros, we can inspire new applications that may be otherwise unachievable with traditionally used, lower-dimensional dark spots, such as those in beams or simple standing waves.

\section{Methods}
3D electric field zeros were created in analytical simulations of ten monochromatic interfering plane waves.
In all simulations, ten random wavevectors (all of the same magnitude $k=\frac{2\pi}{\lambda}$) were generated, and for each, two orthogonal polarisation basis vectors were defined, representing the two electric field degrees of freedom of a plane wave propagating in that direction.
The ten plane waves were then polarised deliberately to destructively interfere and leave a 3D electric field zero at a single confined point, $\mathbf{r}_0$, following the procedure given in \cite{Vernon2022}.
Let $\textrm{e}^{i\mathbf{k}_j\cdot\mathbf{r}}\mathbf{\hat{e}}_{j,1}$ and $\textrm{e}^{i\mathbf{k}_j\cdot\mathbf{r}}\mathbf{\hat{e}}_{j,2}$ be the two orthogonal polarisation states (degrees of freedom) of the electric field of the $j^{th}$ plane wave with unit amplitude at the origin ($j$ ranges from 1 to 10, $\mathbf{k}_j$ is the $j^{th}$ plane wave's random wavevector with magnitude $|\mathbf{k}_j|=\frac{2\pi}{\lambda}$, and $\mathbf{\hat{e}}_{j,1}$ and $\mathbf{\hat{e}}_{j,2}$ are two orthogonal unit vectors satisfying $\mathbf{\hat{e}}_{j,1}\cdot\mathbf{\hat{e}}_{j,2}=0$, $\mathbf{\hat{e}}_{j,1}\cdot\mathbf{k}_j=0$, $\mathbf{\hat{e}}_{j,2}\cdot\mathbf{k}_j=0$).
In total, we have twenty available polarisation degrees of freedom, and by propagating each plane wave, we can calculate the electric field that each individual degree of freedom develops in the position of a desired electric field zero, $\mathbf{r}_0$. Now, we multiply each degree of freedom by a complex scalar, so that the $j^{th}$ plane wave has components $x_{j,1}\textrm{e}^{i\mathbf{k}_j\cdot\mathbf{r}}\mathbf{\hat{e}}_{j,1}$ and $x_{j,2}\textrm{e}^{i\mathbf{k}_j\cdot\mathbf{r}}\mathbf{\hat{e}}_{j,2}$.
Adding together all scaled degrees of freedom, evaluated at $\mathbf{r=r}_0$, we have a linear system of three equations, one per component of the total field at $\mathbf{r}_0$, with complex variables $x_{j,1}$ and $x_{j,2}$ representing the amplitude of the orthogonal components of the $j^{th}$ plane wave phasor.
Setting to zero all three total electric field components at $\mathbf{r}_0$, we may solve the system of equations to find the polarisation components of each plane wave required for complete destructive interference at $\mathbf{r}_0$.
Since only three scalar conditions are enforced ($E_x=0$, $E_y=0$ and $E_z=0$ for the total field at $\mathbf{r}_0$) by twenty degrees of freedom, the system is under-determined and seventeen possible solutions exist for a 3D zero at $\mathbf{r}_0$.
Any one of these solutions may be chosen to realise the zero, or, as we do, the solutions may be combined in a linear sum with random complex amplitudes.
A 3D zero could be produced with as few as four plane waves (in fact, a zero could be enforced by only two plane waves, but it would not be three-dimensional), though the total field would appear less random.

\printbibliography[title={References}]

\section{Acknowledgements}
We would like to thank Sinuh\'e Perea-Puente for a mathematical proof. 
This work was supported by European Research Council Starting Grant ERC2016-STG-714151-PSINFONI.
\section{Author Contribution}
A.J.V. conducted mathematical analyses and simulations; M.R.D. gave direction to and supervised the research; F.J.R-F. supervised the research. All authors wrote the manuscript; A.J.V. wrote the first draft.
\section{Competing Interests}
The Authors declare no competing interests.
\end{document}